\documentstyle[aps,pra]{revtex}
       \begin{document}
\title{Bosonic and fermionic behavior in gravitational configurations} 
\author{
Leonardo Pati\~no 
\thanks{E-mail address: {\tt leonardo@ft.ifisicacu.unam.mx}} 
and Hernando Quevedo
\thanks{ E-mail address: {\tt quevedo@nuclecu.unam.mx}}} 
\address{
{\it Instituto de Ciencias Nucleares, Universidad Nacional 
Aut\'onoma de M\'exico}\\
A. P. 70-543,  M\'exico 04510 D.F., M\'exico}
\date{\today}
\maketitle
       \begin{abstract}
We extend Dirac's approach about the quantization of the electric charge to the 
case of gravitational configurations. The spacetime curvature is used to define 
a phase-like object which allows us to extract information about the behavior 
of the corresponding spacetime. We show that all spacetimes that satisfy certain 
simple symmetry condition and for which the Petrov type is the same whitin a 
specific region, quantization conditions
can be derived that impose constraints on the possible values of the parameters 
entering the respective metrics. As a general result we obtain that for 
the gravitational configurations described by those metrics, the behavior under rotations
can be only of bosonic or fermionic nature.    
   \end{abstract}
\section{Introduction}
It is widely believed that half--integral spin objects (fermions) appear only in quantum
theories and that they have no classical counter--partners. Objects with integral spin 
(bosons),  instead,  are characteristic in both classical and quantum configurations. 
One of the most important properties of these two kind of objects is the way how 
they behave under rotations: Bosons are invariant under a 2$\pi$ rotation of space 
coordinates, while fermions are invariant only under a $4\pi$ rotation.

An argument in favor of the existence of a half--integral fermionic behavior in classical 
theories was given by Finkelstein and Misner \cite{FinMis}. Later on, this conjecture was 
successfully ``tested" by Friedman and Sorkin \cite{Sorkin2} who analyzed the behavior under 
diffeomorphisms of a Schr\"odinger state vector in the configuration space defined by
asymptotically flat 3-dimensional classical metrics. Although this approach is based upon very
natural assumptions about the kinematic behavior of the gravitational state vector, a complete
theory of quantum gravity will be necessary in order to confirm this result.
More recently, Hadley \cite{Hadley3} has shown that even at the level of classical general
relativity a fermionic behavior naturally appears, when geons are treated as models 
for elementary particles. But the first indication that half--integral spins can arise 
as a composite state of integral spin constituents was given originally by Dirac in 
his seminal paper of 1931 \cite{Dirac1}. Indeed, Dirac analyzed the quantum phase acquired 
by a charged particle when dragged around a closed loop in the presence of a magnetic 
monopole. In this case, the charge-monopole system can have half-integral spin, even 
if the electric charge and the magnetic monopole by themselves have integral spin. 
One further important result of this analysis is that it predicts the quantization 
of the electric charge, if the magnetic monopole exists. 

In this work, we will follow Dirac's original idea which we now briefly review. 
The quantum phase acquired by a charged particle while moving within a magnetic field $B$ 
is given by
\begin{equation}
\Phi =e^{{iq}\int _{\gamma}A},          \label{1}
\end{equation}
where $A$ is the vector potential 
for $B$. \textit{i.e.}  $(dA=B)$, $\gamma$ is the path 
followed by the particle and $q$ is its charge. If  $\gamma$ is a closed path, we can 
use Stoke's theorem to get
\begin{equation}
\Phi=e^{{iq}\int _{S}B},               \label{2}
\end{equation}
with $S$ any surface having $\gamma$ as its boundary.
The magnetic field of a magnetic monopole in spherical coordinates is 
$B= g \sin\theta\ d\theta \wedge d\phi$, where $g$ is the magnetic charge. The interesting
feature of this field is that it is a closed two-form which is not exact. Therefore, 
if we have a charged particle immersed in this field, and we drag it along a circle around 
the origin of the coordinate system, we can use Eq.(\ref{2}) to calculate the phase, but we cannot use Eq.(\ref{1}), because there is no such an $A$. In order to calculate the phase (\ref{2}) we can use different surfaces, 
in particular we can use the two different hemispheres which have $\gamma$ as their  boundary, and $\gamma$ is usually taken as a circle on the equatorial plane\footnote{Actually the circle must be vertical, otherwise the points $\theta =0$ and $\theta = \pi$ , which are in the boundary of the domain of integration, get mapped into the interior of the surface over the manifold.}.
The sign of the integral over one hemisphere must be the opposite of the sign of the integral over the other hemisphere in order to be consistent with the orientation of  $\gamma$. Then we get $\Phi=\exp(\frac{-iqg}{2} )$ for one hemisphere and $\Phi=\exp(\frac{iqg}{2} )$ for the other one. If we demand that these two results be equal, we get  that  $\frac{qg}{2} =n \pi$, for  any integer $n$. 
Dirac concluded from here that the existence of a magnetic monopole will imply the 
quantization of the electric charge $q=n \frac{ 2 \pi }{g}$.

In this work, we will apply Dirac's approach to the case of gravity. In 
Section 2, we introduce a phase-like object which explicitly contains  a surface 
integral of the spacetime curvature. A brief discussion about some details of this 
definition is included. Section 3 is devoted to the investigation of the symmetries of the eigenvalues of the
curvature tensor. We also introduce the symmetry of spacetime that is necessary
in order to perform a quite general calculation of the phase-like object which is
the subject of Section 4. In this Section, we also impose Dirac's quantization 
condition on our phase-like object and prove that it leads to the conclusion that 
classical gravitational configurations may behave either as bosons or as fermions
under rotations and that, in some special cases, the parameters entering
the spacetime metric have to satisfy certain quantum conditions. 
Finally, in Section 6 we present the conclusions of our results.

\section{Generalization to gravitation}
First, let us notice that the integral in Eq.(\ref{2}) is in fact an integral
of the electromagnetic tensor over a specific surface. In the special case 
of a magnetic monopole, the only non vanishing entrance of the electromagnetic tensor is
the one corresponding to $d\theta \wedge d\phi$ in spherical coordinates.
On the other hand, from the geometrical point of view of field theory the electromagnetic
tensor is just the curvature associated to the electromagnetic connection. 
Consider a Riemannian manifold $(M,\ g_{\mu \nu})$, where  $g_{\mu \nu}\ 
(\mu, \ \nu=0,1,2,3)$ is the underlying metric. Then we can introduce  
a phase-like object\footnote{An exact definition would need, in general, 
 a path-ordered exponential and a constant factor in front of the integral depending 
on the choice of units.}
\begin{equation}
\Phi=e^{\int R},   \label{3}
\end{equation}
where $R$ is the curvature associated with the metric, namely the Riemann tensor, 
which is an endomorphism valued two-form (see, for instance, \cite{Baez}). 
The integral in Eq.(\ref{3}) has to be performed over the two-form part of 
the Riemann tensor, and the result is an endomorphism. So, 
the phase-like object (\ref{3}), 
being the exponential of an endomorphism, is itself an endomorphism.

The Riemann tensor depends on the point of the manifold where it is evaluated, since 
it is a two-form which components are endomorphisms. These components actually live in the tangent and cotangent spaces to the manifold at the specific point of evaluation. 
If we perform the integral in Eq.(\ref{3}) just as it appears, we will be adding objects 
which live in different spaces and this summation will not be justified. 
What we need is to perform a parallel translation of the endomorphism\footnote{This parallel translation has to be performed only over the endomorphism part of the Riemann tensor;
 the two-form will be handled using the pull-back in the conventional way.} from the point of evaluation to a specific point where we will say that the integral is based.
So, instead of (\ref{3}) we should use  
\begin{equation}
\Phi=e^{\int H^{-1}R H},  \label{4}
\end{equation}
where $H$ is the holonomy resulting from the parallel transport,
 and it will depend explicitly on the point of evaluation.
Of course, there are infinite many different paths along which the parallel transport can 
be performed. Therefore, the specific form of $H$ will have to be fixed by using
a different approach \cite{lh}. For the moment, the important point is that there are some conclusions that can be stated independently of the explicit form of the holonomy $H$.
The only condition we will require below is that it preserves some fundamental symmetries of the curvature. Since $H$ is directly related  to the metric and its symmetries, it seems 
natural to expect this behavior. Otherwise, it can be imposed by construction.

The key point of the choice of surfaces in Dirac's argument is that the two hemispheres are not homotopic. So the physical relevant cases expected in the analysis of gravitational configurations
will be those topological spaces which are solutions to the Einstein equations and 
accept the existence of not homotopic surfaces with common boundary, for instance,
manifolds with localized curvature singularities.

\section{Symmetries}

In order to analyze the phase-like object defined in the previous section for 
gravitational fields in a general fashion, we will consider the 
curvature tensor in its $SO(3,C)$-representation~\cite{Petrov}. Moreover, 
we will limit ourselves 
to \textit{vacuum} solutions of Einstein's field equations\footnote{For non 
vacuum gravitational fields, our approach can be applied with no changes to the
Weyl tensor.}. The curvature tensor
is represented by a $3\times 3$ complex matrix ${\bf Q}$, the entrances of which are complex
combinations of the components of the Riemann tensor written in a local orthonormal
tetrad $\{e^a\}$, with local metric $\eta_{ab}=$diag$(-1,\ 1,\ 1,\ 1)$.
 According to the Petrov classification, there exists a
special orthonormal tetrad in which the matrix ${\bf Q}$ takes a very simple form, the
\textit{normal} form. We will see that this normal form allows us to 
extract certain symmetries of the curvature which are crucial for the calculation of
the phase-like object (\ref{4}). 

Consider, for instance, the normal form of a type $I$ curvature tensor~\cite{Petrov}
\begin{equation}
{\bf Q}= \left( \begin{array}{ccc}
 \tau _{1} & 0 & 0 \\
 0 & \tau _{2} & 0 \\
 0 & 0 & \tau _{3}
\end{array} \right), \label{tipoI}
\end{equation}
where $\tau _{n}\ (n=1,2,3)$ are the eigenvalues of the curvature,
which are subject to the constraint  $\tau _{1}+ \tau _{2}+ \tau _{3}=0$,
 imposed by
Einstein's vacuum field equations. For concrete calculations of the integral in 
Eq.(\ref{4}), we need to know the explicit components of the curvature tensor. 
These components can be recovered from the normal form (\ref{tipoI}). For example,
the components associated with the two-form $e^1\wedge e^2$ can be expressed as the
$4\times 4$ real matrix\footnote{We use the following convention: ${\bf Q} =
{\bf A} + i{\bf B}$, where ${\bf A}$ and ${\bf B}$ are real $3\times 3$ matrices 
which determine the $6\times 6$ curvature matrix
\begin{equation}
{\bf R}= \left( \begin{array}{cc}
 {\bf A}& {\bf B} \\
 {\bf B} & -{\bf A}
\end{array} \right).
\end{equation}
All the independent components of the curvature tensor are contained in the matrix ${\bf R}$. 
Our convention for the transition between the 
matrix entrances of ${\bf R}$ and the pair of tetrad indices is~\cite{mtw}: 
$1\rightarrow 01$, $2\rightarrow 02$,
$3\rightarrow 03$, $4\rightarrow 23$, $5\rightarrow 31$, $6\rightarrow 12$.}
\begin{equation}
R^{a}_{\ b12}= \left( \begin{array}{cccc}
0 &0 &0 &-Im(\tau_{3}) \\
0 &0 &-Re(\tau_{3}) &0 \\
 0& Re(\tau_{3}) &0 &0 \\
 -Im(\tau_{3}) &0 &0 &0
\end{array} \right), \label{r12}
\end{equation}
where $Re$ and $Im$ stand for the real and imaginary part of the argument, respectively. 
Now, since the Riemann tensor is an endomorphism valued two-form, when we talk about its 
components, we refer to the endomorphisms associated with each entrance of the two-form. 
These components map a four dimensional space into itself, so they must have four 
eigenvalues $\lambda _{i}$. In the explicit example of Eq.(\ref{r12}), these eigenvalues 
are given by
$\lambda_{1}=-Im(\tau_{3}), \; \lambda_{2}=Im(\tau_{3}), \; \lambda_{3}=iRe(\tau_{3}),$ 
and $ \lambda_{4}=-iRe(\tau_{3})$, and satisfy the relationships
\begin{equation}
\lambda _{1}=-\lambda _{2}  \ \ \ \ \ \ {\rm and} \ \ \ \ \ \  
\lambda _{3}=-\lambda _{4}. \label{sym}
\end{equation}
Now, the important fact is that Eq.(\ref{sym}) holds for all the components of the
curvature tensor. The explicit form of the matrix (\ref{r12}) may be different, but 
in all the cases the eigenvalues satisfy the symmetry property (\ref{sym}). 
Furthermore, it can be shown that this property is also valid for all the algebraically 
special curvature tensors, \textit{i.e.}, for the Petrov types $D, \ II, \ N, \ III,$ and
$O$. That is to say, the relationships (\ref{sym}) constitute a universal property of 
vacuum gravitational fields. The importance of this result for the calculation
of the phase (\ref{4}) will be demonstrated in the next section.

The invariant character of the symmetry (\ref{sym}) follows from the fact 
that the Weyl principal tetrad $\{e^a\}$, in which the curvature tensor
takes its normal form, is uniquely determined (up to trivial reflections)
for the non degenerate Petrov types ($I,\ II,$ and $III)$. In the case 
of the degenerate types ($D, \ N,$ and $O$), the Weyl principal tetrad is
fixed only up to  special Lorentz transformations~\cite{Petrov} which, however, 
do not affect the invariant character of the curvature eigenvalues $\tau_n$ and,
consequently, preserve the symmetry property (\ref{sym}). Moreover, it is diffeomorphism
invariant because changes of coordinates do not affect the local tetrads. 
In the practice, a local change of coordinates implies a summation over a specific linear 
combination of the components of the curvature tensor. From the form of the
Riemann tensor, as extracted from its normal form, it is not difficult to show that any 
linear combination of its components has eigenvalues satisfying Eq.(\ref{sym}). 

The explicit calculation of (\ref{4}) includes the holonomy $H$. Therefore, we have
to guarantee that it does not affect the symmetry property (\ref{sym}). To do this,
we first have to demand that the type in the Petrov classification of the curvature
tensor does not change on any point of the path defined by $H$. Hence, we will 
limit ourselves to those regions of spacetime in which no changes of the Petrov type 
occur. This seems to be sufficient for satisfying Eq.(\ref{sym}) on all the points
of the path, but since in this work we are not giving an exact definition
of $H$, we prefer for the moment 
to impose the further condition that the holonomy $H$ behaves in such a 
way that the eigenvalues of the endomorphism resulting from the integration 
 satisfy Eq.(\ref{sym}).

Let us now describe the surfaces of integration. 
We can construct a closed surface symmetric with respect to an axis, say $\mathbf{z}$,
and intersect it with any plane which contains this axis. In this way,  we obtain two closed surfaces with the same boundary. If the spacetime is described 
by a Riemannian metric with a localized curvature 
singularity which is placed between the two surfaces, then the two surfaces are non homotopic.
We now introduce spherical coordinates. Let us suppose that
the spacetime metric is invariant with respect to the coordinate 
change\footnote{Clearly, this condition is satisfied by a large
class of gravitational configurations, for instance, all the axially symmetric 
solutions.}
$\phi \longrightarrow \phi + \pi$. Now, by virtue of this symmetry, 
the integrals of the Riemann tensor over the two surfaces can differ only by a 
sign\footnote{The difference in the sign comes from the fact  that when the metric is 
symmetric with respect to some transformation, the Riemann tensor can be either symmetric or antisymmetric with respect to it.}. We will denote these two integrals by  
$I _{1}$ and $I _{2}$, and the corresponding phases by $\Phi_1=$exp$(\pm I_1)$ and 
$\Phi_2=$exp$(\mp I_2)$, where the sign has to be chosen in accordance with the
orientation of the boundary.

\section{Condition of ``quantization"}

Following Dirac's argument we demand that the phase-like object (\ref{4}) has the same
value on both surfaces, \textit{i. e.}, $\Phi_1=\Phi_2$, a requirement that we call 
condition of ``quantization". If the integrals $I _{1}$ and $I _{2}$ have opposite signs, 
then the phase-like objects $\Phi _{1}$ and $\Phi_{2}$ obtained from them are equal, 
since to be consistent with the orientation of the boundary one of the signs must be 
reversed before exponentiating. In this situation the requirement $\Phi_{1}= \Phi_{2}$ is 
trivially satisfied, and no additional condition can be obtained from the analysis. 

On the other hand, if  $I _{1} = I _{2}$, then due to the change of sign that has to be done
in one of these two integrals, say $-I_2$,  before exponentiating, $\Phi_{1}$ and 
$\Phi_{2}$ will be different and, 
consequently, a further analysis is required. 
In this case, the eigenvalues of  $I_{1}$ differ from the eigenvalues of $-I_{2}$ only 
by a minus sign. This together with the symmetry property (\ref{sym}) implies that 
both sets of eigenvalues coincide and the only difference appears in the interchange 
of the eigenvectors associated with each of these eigenvalues. Now we have to exponentiate
the integrals $I$ which are in general $4\times 4$ matrices following from the integration
of each of the entrances of the Riemann tensor as given, for example, in Eq.(\ref{r12}).

A simple way to deal with the exponential of a matrix, say $M$, is first to diagonalize it
and then to calculate the exponential as
\begin{equation}
\exp(M)=T \, \exp(D) \, T^{-1},
\end{equation}
where $D$ is the diagonal matrix which has the eigenvalues of $M$ as components, and the matrix
$T$ represents the change of basis and is constructed explicitly by placing the eigenvectors of  $M$ as columns in the same order as the associated eigenvalues appear in $D$. 
Following this procedure for $\Phi_{1}$ and $\Phi_{2}$, we see that  
since $I_{1}$ and $-I_{2}$ have the same eigenvalues $\lambda _{i}$ satisfying the 
symmetry property (\ref{sym}), we can define a matrix 
\begin{equation}
A \equiv \exp \left( \begin{array}{cccc}
\lambda_{1} & 0 & 0 & 0 \\
0 & -\lambda _{1} & 0 & 0 \\
0 & 0 & \lambda _{3} & 0 \\
0 & 0 & 0 & -\lambda _{3}
\end{array} \right) = \left( \begin{array}{cccc}
e^{\lambda _{1}} & 0 & 0 & 0 \\
0 & e^{-\lambda _{1}} & 0 & 0 \\
0 & 0 & e^{\lambda _{3}} & 0 \\
0 & 0 & 0 & e^{-\lambda _{3}}
\end{array} \right),      \label{A}
\end{equation}
and then express
\begin{equation}
\Phi_{1}=T_{1}AT_{1}^{-1} \; , \ \ \ \; \Phi_{2}=T_{2}AT_{2}^{-1}.      \label{9}
\end{equation}
From Eq.(\ref{9}) it follows that if we impose the condition of quantization 
\begin{equation}
\Phi_{1}=\Phi_{2} \; \Leftrightarrow \; [A,T_{2}^{-1}T_{1}]=0,     \label{10}
\end{equation}
where square brackets denote the commutator of the corresponding matrices. Thus, 
the verification of the vanishing of this commutator is a very easy way to extract
information from the quantization condition $\Phi_1=\Phi_2$.

As we have mentioned above, the eigenvectors of  $I_{1}$ and $I_{2}$ are the same; 
the only differences is that the associated eigenvalues are interchanged. Therefore, 
the matrices $T_{1}$ and $T_{2}$ have the same columns in different order. 
Then, the general result is
\begin{equation}
T_{2}^{-1}T_{1}=\left( \begin{array}{cccc}
0 & 1 & 0 & 0\\
1 & 0 & 0 & 0\\
0 & 0 & 0 & 1\\
0 & 0 & 1 & 0
\end{array} \right).
\end{equation}
From here and Eq.(\ref{A}) we see that the commutator in Eq.(\ref{10}) will vanish
only if 
\begin{equation}
e^{\lambda_{1}}= e^{-\lambda_{1}}$ \ \ and \ \ 
$ e^{\lambda_{3}}= e^{-\lambda_{3}},    \label{11}
\end{equation}
an equation that leads to the conditions
\begin{equation}
\lambda_{1}=i \, n_{1} \, \pi \; \; \; $ \ \ and \ \ $ \;\;\; \lambda_{3}=i \, n_{2} \, \pi , \label{CC}
\end{equation}
where $n_1$ and $n_2$ are arbitrary integers. 
Remember that the eigenvalues $\lambda_{i}$ depend on the parameters of the metric. Then, 
Eqs.(\ref{CC}) can be interpreted as quantum conditions on those parameters\footnote{It is true as well that in general the $\lambda$'s depend on the coordinates which were not integrated in (\ref{4}), but the independence of these coordinates could be required as an additional condition which would further specify the metrics that 
would lead to ``realistic" quantum conditions.}, similar to that obtained by Dirac.
Some concern can be risen by the imaginary character of the $\lambda$'s imposed by Eqs.(\ref{CC}); however, in explicit calculations these eigenvalues arise always as square 
roots of general functions which may have positive and negative values, depending on the parameters of the metric, giving place to real and imaginary $\lambda$'s. For instance, 
in our example of Eq.(\ref{r12}) we have that $\lambda_1$ is real, but $\lambda_3$ is
imaginary. 
When an eigenvalue is real, the only solution to Eqs.(\ref{CC}) is that its associated $n$ is equal to zero, but when the eigenvalue is imaginary the $n$ can be any integer, leading
to an infinite set of discrete solutions. 

Now, let us consider the case in which the quantum conditions are not trivially
satisfied, \textit{i. e.}, Eqs.(\ref{CC}) hold for $n_1,\ n_2\neq 0$. Then 
by using Eqs.(\ref{CC}) in Eq.(\ref{A}) we obtain for $A$  
four different possibilities 
\begin{equation}
A_{1}= \left( \begin{array}{cc}
1 \! \! \mbox{\large{1}}& \mbox{\large{0}}\\
\mbox{\large{0}} & 1 \! \! \mbox{\large{1}}
\end{array} \right), \: 
A_{2}= \left( \begin{array}{cc}
1 \! \! \mbox{\large{1}}& \mbox{\large{0}}\\
\mbox{\large{0}} & -1 \! \! \mbox{\large{1}}
\end{array} \right), \: 
A_{3}= \left( \begin{array}{cc}
-1 \! \! \mbox{\large{1}}& \mbox{\large{0}}\\
\mbox{\large{0}} & 1 \! \! \mbox{\large{1}}
\end{array} \right), \: \ \  \mathrm{or} \ \  \:  
A_{4}= \left( \begin{array}{cc}
-1 \! \! \mbox{\large{1}}& \mbox{\large{0}}\\
\mbox{\large{0}} & -1 \! \! \mbox{\large{1}}
\end{array} \right), \label{ai}
\end{equation}
where $ \mbox{\small{1}} \! \! 1 $ is the $2\times 2$ unit matrix. Inserting these
expressions into Eq.(\ref{9}), we obtain a phase-like object which is either the identity\footnote{Since we have already 
imposed the condition $\Phi_{1}=\Phi_{2}$, we can use indistinctly $T_{1}$ or $T_{2}$ for the following calculations.}
\begin{equation}
\Phi=TA_{1} T^{-1}= \mbox{\small{1}} \! \! 1_{4 \times 4}, \label{bos}
\end{equation}
or something different from the identity for $A_k\ (k=2,3,4)$ 
\begin{equation}
\Phi=TA_{k} T^{-1}, \label{ferm1}
\end{equation}
an expression whose square, however, becomes the identity, \textit{i. e.},
\begin{equation}
\Phi^{2}=TA_{k} T^{-1}TA_{k} T^{-1}=\mbox{\small{1}} \! \! 1_{4 \times 4}, \label{ferm2}
\end{equation}
where we have used the fact that $A_{i}^{2}=\mbox{\small{1}} \! \! 1_{4 \times 4}$, as can 
easily be verified from Eq.(\ref{ai}). 

This is an interesting result, because when understood as an active diffeomorphism, 
the action of an observer traveling along a loop around an object is  equivalent to 
rotating that object by $2\pi$. 
Different authors have tried to model elementary particles by means of exact solutions 
of Einstein's equations~\cite{Einstein,Wheeler,Sorkin1,Sorkin2,Balach2}. If we assume that 
point of view here, we see that a particle described by a metric with 
the symmetry properties discussed above behaves under rotations only in two possible ways, 
as a boson in the case specified by Eq.(\ref{bos}), or as a fermion in the case specified 
by Eq.(\ref{ferm1}) and Eq.(\ref{ferm2}). Several investigations have been devoted 
to the search of this kind of behavior in models of elementary 
particles~\cite{Sorkin1,Hadley3}. In this work we have found  this behavior 
in gravitational configurations by using a very simple approach.

\section{Concluding Remarks}

We have shown that Dirac's argument about charge quantization in the presence of a magnetic monopole can be reproduced in the context of gravitation. We have defined a phase-like 
object in Eq.(\ref{4}) that allows us to extract information about the behavior under rotations of any object enclosed within the surface of integration. 
In this direction further work has to be done with respect to the definition of the integral of the Riemann tensor over surfaces as an endomorphism valued two-form subject to a parallel transport. Although this definition is not given explicitly in the present work, some general conclusions have been stated, assuming that the symmetries of the system are respected by the final form of the integral. 

We have used the normal form of the curvature tensor and its Petrov classification to show
an invariant symmetry property given in Eq.(\ref{sym}) of the eigenvalues of a matrix which is constructed by a set of specific components of the curvature tensor, and enters the definition of the phase-like object (\ref{4}). We then demand that the metrics under consideration be 
invariant with respect to a rotation by $\pi$ of the azimuthal angle. This condition and 
the symmetry property (\ref{sym}) give enough information to calculate in terms of eigenvalues and eigenvectors the phase-like object $\Phi$ on each of the surfaces of integration, assuming
that the the matrix integral $I$ does not affect these symmetries.  By 
imposing the quantization condition $\Phi_1 = \Phi_2$, we obtain an expression 
relating only the eigenvalues of the matrix integrals $I_1$ and $I_2$. When the 
integration surfaces are homotopic, the quantization condition turns out to be 
trivially satisfied, whereas in the case of non homotopic surfaces we obtain 
conditions on the eigenvalues of the curvature tensor. These conditions indicate 
that the parameters entering the metric cannot be arbitrary, but they must be related by
a discrete set of integer values. Additionally, we have shown that spacetimes 
satisfying the corresponding symmetry condition behave under rotations either as bosons or 
as fermions. First, it is interesting to see that gravitational configurations show this characteristic fermionic behavior even at the classical level. On the other hand, a quite 
remarkable result of our analysis is that in classical gravitational configurations 
no other possibilities are allowed, apart from bosonic or fermionic behavior. 

Of course, to construct a realistic model of elementary particles many other aspects
have to be considered, for instance, the requirement of the bosonic or fermionic behavior 
under the interchange of  particles, the spin-statistics theorem~\cite{Aneziris,Aneziris2,Balach4}, etc. Nevertheless, it is interesting to see that 
under this perspective one of these aspects is predicted correctly by the approach 
presented in the present work.

\end{document}